\documentclass[aps,prb,twocolumn,superscriptaddress,floatfix]{revtex4}
\usepackage{amsmath,graphicx,bm}
\begin{document}


\title{Exciting half-integer charges in a quantum point contact}


\author{I. Snyman}
\affiliation{Instituut-Lorentz, Universiteit Leiden, P.O. Box 9506, 2300 RA Leiden, The Netherlands}
\author{Y.V. Nazarov}
\affiliation{Kavli Institute of Nanoscience, Delft University of Technology, 2628 CJ Delft, The Netherlands}
\date{\today}

\begin{abstract}
We study a voltage-driven quantum point contact (QPC) strongly coupled to a qubit. We predict pronounced observable features 
in the QPC current that can be interpreted in terms of half-integer charge transfers. Our analysis is based on the Keldysh 
generating functional approach and contains general results, valid for all coherent conductors.
\end{abstract}
\maketitle


The quantum point contact \cite{vWee88}
has become a basic concept in the field of Quantum Transport
owing to its simplicity.
Its common experimental realization
is a narrow constriction that connects two metallic
reservoirs. An adequate theoretical description for this setup is 
a non-interacting one-dimensional electron gas 
interrupted by a potential barrier. 
The barrier is completely characterized by its scattering matrix.
This enables the scattering approach to Quantum Transport \cite{But90}.
This allows one to describe the average current through 
the QPC, as well as fluctuations away from this average, in terms of single electrons passing through the constriction \cite{Lev96}.
The strength of the scattering approach is its ability
to describe not only traditional realizations of a QPC, but 
all coherent conductors, including diffusive wires and tunneling junctions.  

Despite the correctness of the 
non-interacting electron description, 
truly many-body quantum correlations
do exist and are observable in a QPC. 
They manifest themselves in the full counting 
statistics of electron transfers \cite{Lev96}  
and allow for detection of two-particle entanglement \cite{Been03}
through the measurement of non-local current correlations. 
This suggests that the observation of many-body
effects in a QPC crucially relies on a proper detection scheme.
In this Letter, we give an example of how
an appropriate detector uncovers such non-trivial 
many-body effects as half-integer charges.

We probe the QPC with a charge qubit. Such a device has already been realized using single and double quantum dots.
Previously, the QPC has been used as a detector of the qubit state \cite{Elz03,Pet04}. We propose a scheme in which the
roles are reversed. Provided the qubit and QPC are
coupled strongly, the switching between the qubit states 
is accompanied by severe Fermi-Sea shake-up in the QPC.
The d.c. current in the QPC is sensitive to the ratio
of the qubit switching rates and thereby provides
information about these severe shake-ups.

Before analising the system in detail, the following qualitative conclusions can be drawn.
The qubit owes its detection capabilities to the following fact: 
In order to be excited it has to 
absorb a quantum $\varepsilon$ of energy from the QPC.
Here $\varepsilon$ is the qubit level splitting, 
a parameter that can be tuned easily in an experiment by
means of a gate voltage.
The QPC supplies the energy
by transfering charge from the high voltage reservoir 
to the low voltage reservoir. The transfer of charge
$q$ allows qubit transitions for level splittings $\varepsilon<qV$,
 $V$ being the bias voltage applied.

We can assume that successive switchings of the qubit
between its states $\left|1\right>$ and $\left|2\right>$ 
are rare and uncorellated.
The qubit dynamics are then characterized by
the rates $\Gamma_{21}$ to switch
from state $\left|1\right>$ to state $\left|2\right>$ and
$\Gamma_{12}$ from $\left|2\right>$ to $\left|1\right>$. 
The stationary probability to find the qubit in state 
$\left|2\right>$ 
is determined by detailed balance to be
$p_2=\Gamma_{21}/(\Gamma_{12}+\Gamma_{21})$. 
This probability can be observed experimentally
by measuring the current in the QPC. 
The current displays random telegraph noise, 
switching between two
values $I_1$ and $I_2$. 
These correspond to the qubit being in the 
state $\left|1\right>$ or $\left|2\right>$ respectively. The 
d.c. current $I$ gives the average over many switches and is
thus related to the stationary probability by $I=(1-p_2)I_1+p_2I_2$. 
The values of $I_1$, $I_2$ and $I$ are determined through measurement
and $p_2$ is inferred.

When the QPC and qubit are weakly coupled \cite{Ale97,Levi97}, 
a single electron is transfered \cite{Onac06}. 
This liberates at most energy $eV$, implying that the 
rate $\Gamma_{21}$ is zero when $\varepsilon>eV$
and the rate $\Gamma_{12}$ is zero when $\varepsilon<-eV$.
The resulting $p_2$ changes from $1$ to $0$ upon increasing $\varepsilon$
within the interval $-eV<\varepsilon<eV$.
Cusps at $\varepsilon=\pm eV$ signify that charge $e$
is transferred. [See Fig. (2{\bf a})]

Guided by our understanding of weak coupling we can speculate as follows about
what happens at strong coupling. 
Apart from single electron transfers, 
we also expect the coordinated transfers of groups of electrons.
A group of $n$ electrons can provide up to $neV$ of energy
to the qubit.
Therefore, peculiarities in $p_2$ should appear at the corresponding
level splittings $\varepsilon=\pm n eV$, $n=1,2,3,\ldots$\cite{Tob06}
However, it is not apriori obvious that these
peculiarities are pronounced enough to be observed.
The reason is the decoherence of the qubit states 
induced by electrons passing through the QPC. 
The Fourier transform of the qubit transition rate
acquires an exponential damping factor 
$e^{-W|t|}$,  $W^{-1}$ being the decoherence time.
This smoothes out peculiarities at the energy scale $W$.
In the strong coupling regime, the decoherence time is estimated to be
short, $W \simeq eV/$. As a result, it is not clear whether peculiarities at $neV$
are the dominant feature at strong coupling.

Therefore, strong coupling of the QPC and the qubit requires quantitative analysis.
We have reduced the problem to the evaluation of a determinant
of an infinite-dimensional Wiener-Hopf operator.
We calculated the determinant numerically and found that
peculiarities at multiples of $eV$ are minute.
Their contribution to $p_2$ does not exceed $10^{-4}$ and is 
seen only at logarithmic scale and at moderate couplings.
Instead,  far more prominent features occurs at $\varepsilon=\tfrac{1}{2} eV$. 
General reasoning does not predict this.
Straight-forward energy balance arguments force us to conclude that qubit switching is accompanied
by the transfer of charge $e/2$ through the QPC. 
This frees up 
energy $eV/2$, stimulating qubit transitions when $\varepsilon<eV/2$. 
In other words, the qubit switching excites 
a half-integer charge and simultaneously detects it.
Fractional charge is known to occur in strongly 
interacting many-electron systems \cite{Lau83,Jac76,Sut90}
in equilibrium. In contrast to this, 
the electrons in the QPC can be regarded non-interacting
except during the short time the qubit is switching. 
Our system is also unusual in that the half-integer charge 
is only produced during qubit switching and is not present in
the equilibrium state.

\begin{figure}[b]
\begin{minipage}{\columnwidth}
\resizebox{.95 \columnwidth}{!}{\includegraphics{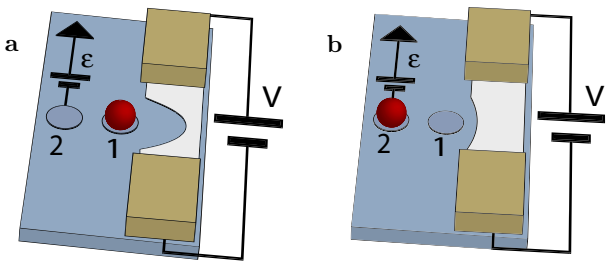}}
\caption{A schematic picture of the system considered. It consists of a charge qubit
coupled to a QPC. The shape of the QPC constriction, and hence its scattering matrix,
depends on the state of the qubit. The QPC is biased at voltage $V$. A gate voltage controls
the qubit level splitting $\varepsilon$. There is a small tunneling rate $\gamma$ between 
qubit states.\label{fig1}}
\end{minipage}
\end{figure}

Let us now turn to the details of our analysis.
The system is illustrated in Fig.\ (\ref{fig1}).
The Hamiltonian for the system is
\begin{equation}
\hat H=\hat{T}+\hat{U}_1\left|1\right>\left<1\right|+(\hat{U}_2+\varepsilon)\left|2\right>\left<2\right|
+\gamma(\left|1\right>\left<2\right|+\left|2\right>\left<1\right|)\label{Ham}
\end{equation}
The operator $\hat T$ represents the kinetic energy of QPC electrons. 
The operator $\hat{U}_k$ 
describes the potential barrier seen by QPC electrons when the qubit is in state $k=1,\ 2$   
and corresponds to a scattering matrix $\check s_k$ in the scattering approach. (We use a ``check'' to
indicate a matrix in the space of transport channels.)
QPC electrons do not interact directly with each other but rather with the qubit.  
This interaction is the only qubit relaxation mechanism included in our model.
We work in the limit $\gamma \to 0$ 
where the inelastic transition rates $\Gamma_{12,21}$
between qubit states are small
compared to the energies $eV$ and $\varepsilon$.
In this case, the qubit switching events can be regarded as independent and incoherent.

Now consider the qubit transition rate $\Gamma_{21}$. To lowest order in the tunneling amplitude $\gamma$ it is given by
\begin{align}
\Gamma_{21}=&2\gamma^2{\rm Re}\,\int_{-\infty}^0d\tau\,e^{i\varepsilon\tau}\nonumber\\
&\times\lim_{t_0\to-\infty}{\rm tr}\left[e^{i\hat{H}_2\tau}e^{-i\hat{H}_1(\tau-t_0)}\rho_0 e^{i\hat{H}_1(\tau-t_0)}\right].\label{rate}
\end{align}
This is the usual Fermi Golden Rule. The Hamiltonians
$\hat{H}_1$ and $\hat{H}_2$ are given by $\hat{H}_k=\hat{T}+\hat{U}_k$ and represent QPC dynamics when the qubit is held
fixed in state $\left|k\right>$. 
The trace is over QPC states, and $\rho_0$ is the initial QPC density matrix.
The evaluation of the integrand is a special case of a general problem in the extended Keldysh formalism \cite{Naz03}. The task is to evaluate the
trace of a density matrix after ``bra's'' have evolved with a time-dependent Hamiltonian $\hat{H}_-(t)$ and ``kets'' with a different
Hamiltonian $\hat{H}_+(t)$.
\begin{equation}
e^{\mathcal A}={\rm tr}\left[\mathcal T^+e^{-i\int_{-\infty}^\infty dt\,\hat{H}_+(t)}\rho_0 \mathcal T^-
e^{i\int_{-\infty}^\infty dt\,\hat{H}_-(t)}\right].
\end{equation}
We implemented the scattering approach to obtain the general formula 
\begin{equation}
\mathcal A={\rm tr\,ln}\left[\hat{s}_-(1-\hat{f})+\hat{s}_+\hat{f}\right]-{\rm tr\, ln\,}\hat{s}_-.\label{action}
\end{equation}
The operators $\hat s_\pm$ and $\hat f$ have both continuous and discrete indices.
The continuous indices refer to energy, or in the Fourier transformed representation, to time.
The discrete indices refer to transport channel space.
The operators $\hat{s}_{\pm}=\check s_\pm(t)\delta(t-t')$ are diagonal in time. The time-dependent
scattering  matrices $\check s_\pm(t)$ describe scattering by the Hamiltonians $\hat{H}_\pm(t)$ at instant $t$.
(It is the hall-mark of the scattering approach to express quantities in terms of 
scattering matrices rather than Hamiltonians.)
The operator $\hat{f}=\check f(E)\delta(E-E')$ is diagonal in the energy representation.
The matrix $\check{f}(E)$ is diagonal in 
channel space, representing the individual electron filling factors
in the different channels.
A full derivation of Eq. (\ref{action}) will be given elsewhere.
It generalizes similar relations published in \cite{Aba04,Aba05}.

In order to apply the general result to Eq. (\ref{rate}), the 
time-dependent scattering matrices $\check{s}_\pm(t)$ are chosen as 
\begin{align}
\check{s}_+(t)=&\check{s}_1+\theta(t-\tau)\theta(-t)(\check{s}_2-
\check{s}_1),\\
\check{s}_-=&\check{s}_1.
\end{align}
The QPC scattering matrices $\check{s}_{1}(\check{s}_{2})$ 
with the qubit in the state $1(2)$ are the most important
parameters of our approach.

Without a bias-voltage applied,
the QPC-qubit setup exhibits the physics
of the Anderson orthogonality catastrophe \cite{And67}.
For the equilibrium QPC, the problem
can be mapped \cite{Aba04} onto the classic 
Fermi Edge singularity (FES) problem 
\cite{Mah67,Noz69,Mat92}. 
The authors of \cite{Aba04} effectively computed $\mathcal A$ 
in equilibrium. Our setup is simpler than
the generic FES problem since there is no tunneling
from the qubit to the QPC. 
As a result, not all processes considered in \cite{Aba04} are relevant for
our setup. We only need the so-called closed loop contribution. 
The relevant part of the FES 
result for our setup is an anomalous power law 
$\Gamma_{21}^{(0)}(\varepsilon)=\theta(-\varepsilon)\frac{1}{|\varepsilon|}
\left(\frac{|\varepsilon|}{E_{\rm c.o.}}\right)^{\alpha}$ for the equilibrium rate.
Here $E_{\rm c.o.}$
is an upper cutoff energy. 
The anomalous exponent $\alpha$  
is determined by the eigenvalues of $\check{s}_2^\dagger\check{s}_1$ \cite{Yam82} as
$\alpha=\tfrac{1}{4\pi^2}\left|{\rm Tr}\,\ln^2 (\hat{s}_f^\dagger \hat{s}_i)\right|$.
The logarithm is defined on the branch $(-\pi,\pi]$. 
For a one or two channel point contact, $0<\alpha<1$. 

We now give the details of our calculation for the rates
{\em out of equilibrium}.
From Eq. (\ref{rate}) and Eq. (\ref{action}) it follows that  
$\Gamma_{21}(\varepsilon)\propto|\gamma^2|\int_{-\infty}^\infty d\tau\ e^{-i\varepsilon \tau} {\rm Det}\ \hat{Q}^{(V)}(\tau)$.
For positive times $\tau$, the operator $\hat{Q}^{(V)}(\tau)$ is defined as
\cite{Aba04}.
\begin{equation}
\hat{Q}^{(V)}(\tau)=1+(\check{s}_2^{-1}\check{s}_1-1)\hat{\Pi}(\tau)\hat{f}^{(V)}
\end{equation}
while for negative $\tau$, $\hat{Q}^{(V)}(\tau)=\hat{Q}^{(V)}(-\tau)^\dagger$ 
The time-interval operator $\hat{\Pi}(\tau)=\delta(t-t')\theta(t)\theta(\tau-t)$ is diagonal in time and
acts as the identity operator in channel space for times $t=t'\in[0,\tau]$ and as the zero-operator outside this time-interval.

For the purpose of numerical calculation of the determinant we have to
regularise $\hat{Q}^{(V)}(\tau)$. This is done by multiplying with the inverse of the zero-bias operator to define a
new operator
$\tilde{Q}(\tau)=\left.\hat{Q}^{(0)}(\tau)\right.^{-1}\hat{Q}^{(V)}(\tau)$. Its determinant is
evaluated numerically. 
The rate $\Gamma_{21}(\epsilon)$ at bias voltage $V$ is then expressed as the convolution
$\Gamma_{21}(\varepsilon)=\int \tfrac{d\varepsilon'}{2\pi}\Gamma^{\rm eq}_{21}(\varepsilon-\varepsilon') \tilde{P}(\varepsilon')$
of the equilibrium rate and the Fourier transform of $\tilde{P}(\tau)={\rm Det}\, \tilde{Q}^{(V)}(\tau)$,
that contains all effects of the bias voltage $V$.

We implemented this calculation numerically, and computed the probability $p_2$ to find the qubit in state $\left|2\right>$. Details of our numerical method are presented in Appendix A.
Our main results are presented in Fig.\ (2). 
\begin{figure}[b]
\begin{minipage}{\columnwidth}
\resizebox{.95 \columnwidth}{!}{\includegraphics{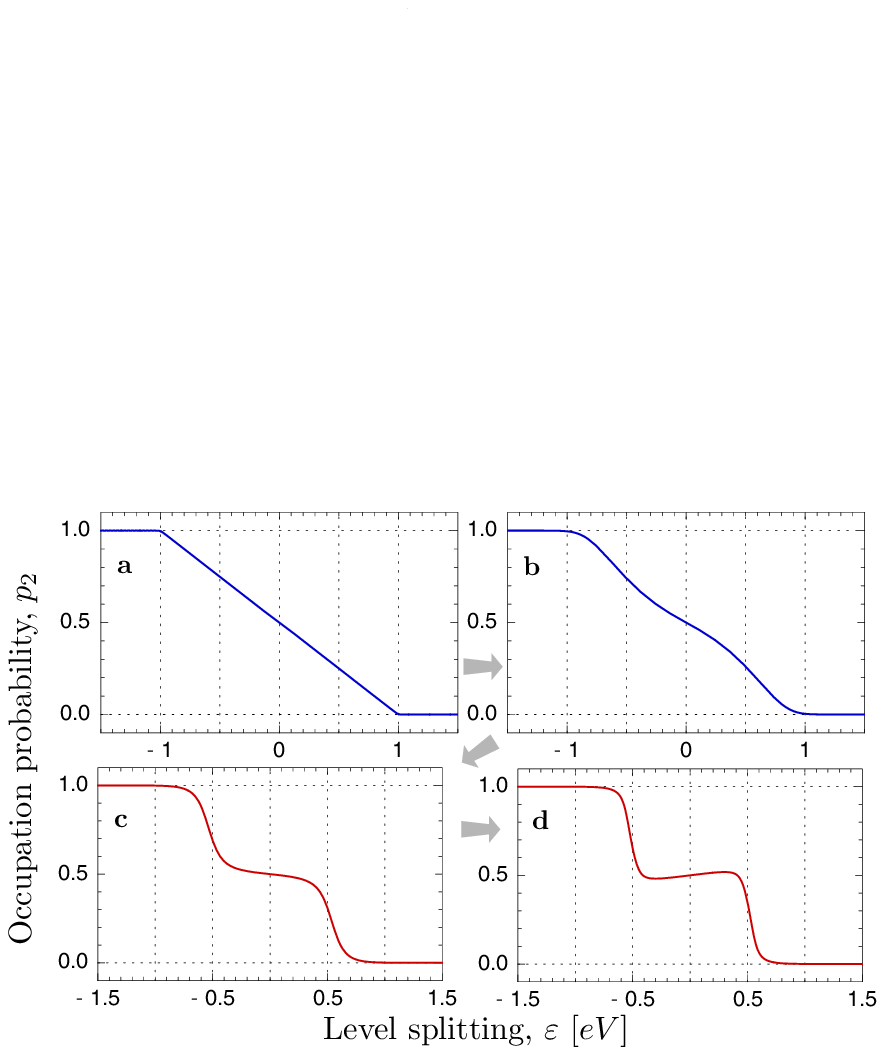}}
\caption{The occupation probability $p_2$ of qubit state $\left|2\right>$. At weak coupling between the
QPC and qubit, (Fig. {\bf a, b}) the transfer of a single electron with charge $e$ is detected. Peculiarities
at $\pm eV/2$ at strong coupling (Fig. {\bf c, d}) constitute the detection of half integer charges $e/2$. Scattering
matrices were parameterized as in Eq. \ref{scat}. Fig. {\bf a, b, c} and {\bf d} respectively correspond to
$\phi=\pi/16$, $\pi/4$, $7\pi/10$ and $4\pi/5$.\label{fig2}}
\end{minipage}
\end{figure}
We used $2\times2$ scattering matrices parametrized by
\begin{equation}
\check{s}_2^{-1}\check{s}_1=\left(\begin{array}{cc} \cos\phi&i\sin\phi\\i\sin\phi&\cos\phi\end{array}\right)\label{scat}
\end{equation}
and repeated the calculation for several $\phi\in[0,\pi]$.
Small $\phi$ corresponds to weak coupling. The 
curve at $\phi=\pi/16$ is almost indistinguishable from the perturbative weak
coupling limit discussed in the introduction.
Cusps at $\pm eV$ indicate that qubit switching is accompanied by the transfer of charge $e$ in the QPC. 

The increasing decoherence smoothes
the cusps for the curve at $\phi=\pi/4$ (2{\bf b}). 
When the coupling is increased beyond $\phi=\pi/2$
steps appear
at $\pm eV/2$ ({\bf c}). 
This implies charge fractionalization $e\rightarrow e/2$.
Further increase of the coupling results in a sharpening of the steps ({\bf d}).


\begin{figure}[b]
\begin{minipage}{\columnwidth}
\resizebox{.95 \columnwidth}{!}{\includegraphics{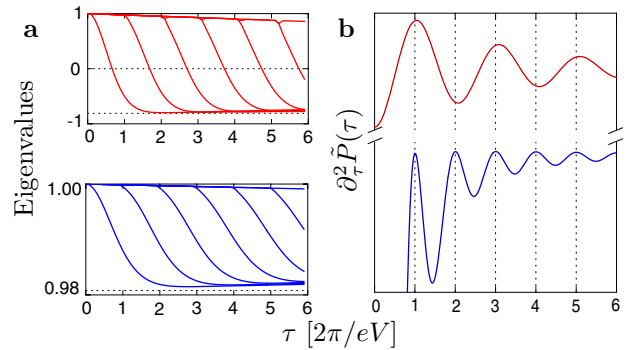}}
\caption{The behavior of eigenvalues for at weak and strong QPC-qubit coupling respectively.
The parameter $\phi$ that parameterises the scattering matrix equals $\pi/16$ (bottom) and $4\pi/5$ (top) representing
the weak and strong coupling limits respectively. 
For $\phi=\pi/16$ individual eigenvalues 
travel from $1$ to $\cos\pi/16\simeq 0.9808$ at a rate of approximately one per $2\pi/eV$. For
$\phi=4\pi/5$, eigenvalues travel towards $\cos 4\pi/5\simeq-0.8090<0$ at a rate of one per $2\pi/eV$, as
shown in ({\bf a}).
Deviations from the correct asymptotics are due to finite size effects. 
Figure ($\bf b$) contains the second derivative of $\tilde{P}(\tau)={\rm Det}\ \hat{Q}^{(0)}(\tau)^{-1}
\hat{Q}^{(V)}(\tau)$. (The second derivative is taken to remove an average slope and curvature.) Oscillations with period $h/eV$ are
seen (bottom) for $\phi=\pi/16$,  while for $\phi=4\pi/5$ (top), the periodicity of $\tilde{P}(\tau)$ doubles.\label{fig3}}
\end{minipage}
\end{figure}

Known mechanisms of charge fractionalization do not 
seem to provide an immediate explanation of our findings.
The Quantum Hall mechanism \cite{Lau83} 
does not give even fractions while
the instanton mechanism \cite{Jac76} requires 
a quasiclassical boson field. There is an indirect analogy
with the model of interacting particles
on a ring threaded by a magnetic flux \cite{Sut90}.
There, one expects that the
energy eigenvalues are periodic in flux with period of one flux quantum.
However, the exact Bethe-Ansatz solution \cite{Sut90} reveals a {\it double} 
period of eigenvalues with adiabatically varying flux. This is a signature of half-integer charge quantization.

For our non-equilibrium setup, energy eigenvalues
are not particulary useful. The natural eigenvalues 
to describe the phenomenon are those of the oprator $\tilde{Q}^{(V)}(\tau)$.
They depend  on the parameter $eV\tau$ which is an analogue of flux.
The product of the eigenvalues, i.e. 
the determinant $\tilde P(\tau)$ is not precisely
periodic in $\tau$ since it decays at large $\tau$ owing to
decoherence. Still, it oscillates and
the period of these oscillations doubles as we go from weak
to strong coupling (Fig. 3{\bf b}). The doubling can be understood
in terms of the transfer of the eigenvalues of $\tilde{Q}^{(V)}(\tau)$
upon increasing $\tau$ (Fig 3{\bf a}) assuming the 
parametrization (\ref{scat}).
In the large $\tau$ limit, 
energy-time uncertainty can be neglected in a ``quasi-classical'' approximation: 
The operator $\hat \Pi(\tau)$ projects onto a very long time interval, and is replaced by the identity operator.
$\tilde Q^{(V)}$ becomes diagonal in energy.
All eigenvalues that are not equal to $1$ are concentrated in
the transport energy window $0<E<eV$
where the filling factors in the QPC reservoirs are not the same. 
 For $\check{s}_2^{-1}\check{s}_1$ parametrized as in (\ref{scat})
these eigenvalues equal $\cos(\phi)$. 
There are $eV\tau/2\pi$ of them.
In other words, the number of eigenvalues equal to $\cos\phi$ grows linearly with $\tau$.
Numerical diagonalization of $\tilde Q^{(V)}(\tau)$ (Fig. 3{\bf a}) shows that
one eigenvalue is transfered from 
$1$ to $\cos(\phi)$ during time $2\pi/ eV$. 
If $\cos\phi >0$ as in the weak coupling case, this gives 
rise to $P(\tau)$ oscillations with frequency $eV/2\pi$ manifesting integer
charges. 
However $\cos\phi$ becomes negative at stronger couplings, 
so that $P(\tau)$ changes sign with each eigenvalue transfer. 
Two eigenvalues have 
to transfer to give the same sign. 
The result is a period {\it doubling}
of the oscillations in $\tilde P(\tau)$ and hence half-integer charges. 
This resembles the 
behavior of the wave vectors of the Bethe-Ansatz solution in
 \cite{Sut90}.

The parametrization (\ref{scat}) of the $\check{s}_2^\dagger \check{s}_1$ is not general. 
However, the eigenvalue transfer arguments help to understand
general scattering matrices. Eigenvalue transfer still occurs at frequency $eV/2\pi$
but instead of traveling along the real line, eigenvalues follow a trajectory inside
the unit circle in the complex plane. Fractional charge is
pronounced if the end point of the trajectory has a negative real part. Numerical results 
for general scattering matrices are presented in Appendix B.

Results presented so far are
for ``spinless'' electrons. Spin degeneracy 
is removed by e.g. high magnetic field. 
If spin is included, but scattering
remains spin independend, 
then two degenerate eigenvalues are transported simultaneously.
In this case, the half-integer charge 
dissapears for the parametrization (\ref{scat})
but persists for the more general choice of complex eigenvalues.
The results of further numerical work that confirm this
are presented in Appendix C.

We have studied a quantum transport setup that can easily 
be realized with current technology, 
namely that of a quantum point contact
coupled to a charge qubit. 
The qubit is operated as a measuring device, 
its output signal --- the probability $p_2$ ---
is directly seen in the QPC current.
The dependence of the signal on 
the qubit level splitting reveals the nature of
charged excitations in the voltage-driven  QPC. 
When the qubit is weakly coupled to the QPC, 
the dependence reveals excitations with electron charge $e$. 
We demonstrated that for stronger coupling,
the dependence suggests the existence 
of the excitations that carry half the charge of an electron.

\appendix
\section{Numerical Method}
In this Appendix we give a more detailed account of the numerical calculation of the qubit tunneling rates 
$\Gamma_{12}(\varepsilon)$ and $\Gamma_{21}(\varepsilon)$ than is presented in the main text.
Our starting point is Eq. (7) of the main text. In order to discuss qubit transitions from 
$\left|1\right>$ to $\left|2\right>$ as well as the reverse transition simultaneously, we change
notation slightly.
In what follows, indices $i$ and $f$ refer to the initial and final state of the qubit respectively. 
We consider ``forward'' transitions $(f,i)=(2,1)$ and ``backward'' transitions $(f,i)=(1,2)$. 
The central object of numerical work is the operator
\begin{equation}
\hat{Q}^{(V)}_{fi}(\tau)=\left\{\begin{array}{lr}
1+(\check{s}^\dagger_i \check{s}_f-1)\hat{\Pi}(-\tau)\hat{f}^{(V)}(\varepsilon)&\tau<0\\
1+(\check{s}^\dagger_f \check{s}_i-1)\hat{\Pi}(\tau)\hat{f}^{(V)}(\varepsilon)&\tau>0
\end{array}\right.
\end{equation}
We recall that the matrices $\check{s}_i$ and $\check{s}_f$ are the
scattering matrices of QPC electrons when the qubit is in state $i$ or $f$. $\hat{\Pi}(\tau)$ is a time-interval operator, 
\begin{equation}
\Pi(\tau)_{t\mu,t'\mu'}=\delta(t-t')\delta_{\mu,\mu'}\left\{\begin{array}{ll}1&\hspace{5mm}0<t<\tau\\
0&{\hspace{5mm}\mbox{otherwise}}\end{array}\right.
\end{equation}
$\hat{f}^{(V)}(\varepsilon)$ is diagonal in
energy. It contains the filling factors of QPC-electrons in the various channels, including any bias voltage that may be present.
Its form in the time-basis (at zero temperature) is given below in Eq.\, (\ref{fdef}). 
The operator $\hat{Q}^{(V)}_{fi}(\tau)$ has an infinite number of eigenvalues outside the neighborhood of 
$1$ in the complex plain. This implies that a regularization of the determinant is needed. 
Indeed, if one naively assumes the unregularized determinant to be well-defined and possesing the usual properties of determinants,
such as ${\rm Det}(AB)={\rm Det}(A){\rm Det}(B)$,
one may show that $\left[{\rm Det}\,\hat{Q}^{(V)}_{fi}(\tau)\right]^*={\rm Det}\,\hat{Q}^{(V)}_{if}(\tau)$. Were this true, it would have implied that 
$\Gamma_{12}(\varepsilon)=\Gamma_{21}(\varepsilon)$. This cannot be correct. At low temperatures, the qubit is far more 
likely to emit energy than
to absorb it, meaning that one of the two rates should dominate the other.

Regularization is achieved by multiplying with the inverse of the equilibrium operator.
The operator $\tilde{Q}_{fi}(\tau)=\hat{Q}_{fi}^{(0)}(\tau)^{-1}\hat{Q}^{(V)}_{fi}(\tau)$
only has a finite number of eigenvalues for finite $\tau$ that are not in the neighborhood of
$1$, and so its determinant can be calculated numerically in a straight-forward manner. 
(In this expression, $\hat{Q}_{fi}^{(0)}(\tau)$ is the operator $\hat{Q}$ when the QPC is initially in
equilibrium, i.e. the bias voltage $V$ is zero.) We therefore proceed as follows: We define
\begin{equation}
\tilde P(\tau)={\rm Det}\left[\hat{Q}_{21}^{\rm (0)}(\tau)^{-1}\hat{Q}^{(V)}_{21}(\tau)\right]
\end{equation}
and $\tilde P(\varepsilon)=\int\,d\tau\,e^{i\varepsilon\tau}\tilde P(\tau)$ as its Fourier transform. The equilibrium rate 
$\Gamma^{\rm eq}_{fi}(\varepsilon)$ is known from the study of the Fermi Edge singularity. It is 
\begin{equation}
\Gamma^{\rm eq}_{fi}(\varepsilon)=|\gamma|^2\theta(-\varepsilon_{fi})\frac{1}{|\varepsilon|}\left|\frac{\varepsilon}{E_{\rm c.o.}}\right|^\alpha
\end{equation} where $E_{\rm c.o.}$ is a cut-off energy of the order of $E_F$ and 
\begin{equation}
\alpha=\frac{1}{4\pi^2}\left|{\rm Tr}\,\ln^2 (\check{s}_f^\dagger \check{s}_i)\right|.
\end{equation}
The logarithm is defined on the branch $(-\pi,\pi]$. With the help of these definitions we have
\begin{equation}
\Gamma_{fi}=\int \frac{d\varepsilon'}{2\pi} \Gamma_{fi}^{\rm eq}(\varepsilon')\tilde P(\varepsilon-\varepsilon'),\label{conv}
\end{equation}
where our task is to calculate $\tilde P(\varepsilon)$ numerically.

The operator $\hat{Q}_{21}^{(V)}(\tau)$ will be considered in the time (i.e. Fourier transform of energy) basis. We restrict
ourselves to the study of single channel QPC's, in which case the scattering matrices $\check{s}_1$ and $\check{s}_2$ are $2\times 2$ matrices in 
QPC-channel space.
We work in the standard channel space basis where 
\begin{equation}
\check{s}_k=\left(\begin{array}{cc} r_k&t'_k\\t_k&r'_k\end{array}\right),
\end{equation}
with $t,\ t'$ the left and right transmission amplitudes and $r,\ r'$ the left and right reflection amplitudes.
Because $\hat{\Pi}(\tau)$ is a projection operator that commutes with the scattering matrices, we can evaluate the determinant
in the space of spinor functions $\psi(t)$ defined on the interval $t\in[0,\tau]$. (We consider $\tau>0$.) Then 
\begin{equation}
\left[\hat{Q}^{(V)}_{21}(\tau)\psi\right](t)=\psi(t)+(\check{s}_2^\dagger \check{s}_1-1)\int_0^\tau dt' \check{f}^{(V)}(t-t')\psi(t')
\end{equation}
where
\begin{eqnarray}
\check{f}^{(V)}(t)&=&\int \frac{d\varepsilon}{2\pi} e^{-i\varepsilon t}\left(\begin{array}{cc}\theta(-\varepsilon)&0\\0&\theta(eV-\varepsilon)
\end{array}\right)\nonumber\\
&=&\frac{i}{2\pi(t+i0^+)}+i\left(\frac{1-\check{\sigma}_z}{2}\right)\frac{e^{-iteV}-1}{2\pi t},\label{fdef}
\end{eqnarray}
is the Fourier transform of the zero-temperature filling factors of the reservoirs connected to the QPC and $0^+$ is an infinitesimal positive constant.
Discretization of this operator proceeds as follows. We choose a timestep $\Delta t\ll \tau$ such that $N=\tau/\Delta t$ is 
a large integer. We will represent $\hat{Q}^{(V)}_{21}(\tau)$ (and $\hat{Q}^{(0)}_{21}(\tau)^{-1}$) as $2N\times2N$ dimensional matrices. We define
a dimensionless quantity $\eta=eV \Delta t$. $\tilde P(\tau)$ can only depend on $\tau$ in the combination $\tau eV$ because
there are no other time- or energy scales in the problem. We will
therefore vary $\tau$ by keeping $N$ fixed and varying $\eta$. Using the identity
\begin{equation}
\frac{1}{t\pm i0^+}={\mathcal P}\left(\frac{1}{t}\right)\mp i\pi\delta(t)
\end{equation}
we find a discretized operator
\begin{eqnarray}
&&\left[1+(\check{s}_2^\dagger \check{s}_1-1)\hat{\Pi}\hat{f}\right]_{kl}\nonumber\\
&&=\delta_{kl}+(\check{s}_2^\dagger \check{s}_1-1)\bigg[\tfrac{1}{2}\delta_{kl} +\frac{1}{2\pi i(l-k)}(1-\delta_{kl})\nonumber\\
&&+\underbrace{\frac{1-\check{\sigma}_z}{2}\left(\frac{\eta}{2\pi}\delta_{kl}
+\frac{e^{i(l-k)\eta}-1}{2\pi i(l-k)}(1-\delta_{kl})\right)}_{\mbox{nonequilibrium 
correction}}\bigg].
\end{eqnarray}
To test the quality of the discretization as well as its range of validity we do the following. When $\check{s}_2^\dagger \check{s}_1$ is close to identity,
we can calculate $\tilde P(\tau)$ perturbatively, both for the original continuous operators and for its discretized approximation.
If we take $\check{s}_2^\dagger \check{s}_1=e^{i\phi\check{\sigma}_x}$ then to order $\phi^2$ we find
\begin{equation}
\tilde P_{\mbox{cont.}}(\tau)=1+2\left(\frac{\phi}{2\pi}\right)^2\int_0^N dz \frac{\cos (z\eta)-1}{z^2}(N-z)
\end{equation}
where $\tau=N\eta/eV$
for the continuous kernel while for the discretized version we find
\begin{equation}
\tilde P_{\mbox{disc.}}(\eta)=1+2\left(\frac{\phi}{2\pi}\right)^2\sum_{\zeta=1}^{N-1}\frac{\cos(\zeta\eta)-1}{\zeta^2}(N-\zeta)
\end{equation}
which indicates that the range of validity is $\eta\ll2\pi$.

In practice we take $N=2^8$. Larger $N$ would demand the diagonalization of matrices that are too large to handle numerically. 
We find results suitably accurate up 
to $\eta=\pi/4$, thereby giving us access to $\tilde P(\tau)$ for $|\tau|\in[0,64\pi/eV]$.

To summerize, the procedure for calculating the transition rates $\Gamma_{21}$ and $\Gamma_{12}$ is
\begin{enumerate}
\item{For given scattering matrices $\check{s}_1$ and $\check{s}_2$, calculate $\tilde P(\tau)$ numerically using the 
discrete approximations for the operators $\hat{Q}^{(V)}_{21}(\tau)$ and $\hat{Q}_{12}^{(0)}(\tau)$.
Use a fixed large matrix size, and work in units $[\tau]=[eV]^{-1}$. Generate data for many positive values of $\tau$.}
\item{Extend the results to negative $\tau$ by exploiting the symmetry $\tilde P(\tau)=\tilde P(-\tau)^*$, and Fourier transform the
data.}
\item{Form the convolutions of Eq.\ \ref{conv} with the known equilibrium rates to obtain the non-equilibrium rates.}
\end{enumerate}
\section{Choice of scattering matrices}
\begin{figure}[b]
\begin{center}
\includegraphics[width=.94 \columnwidth]{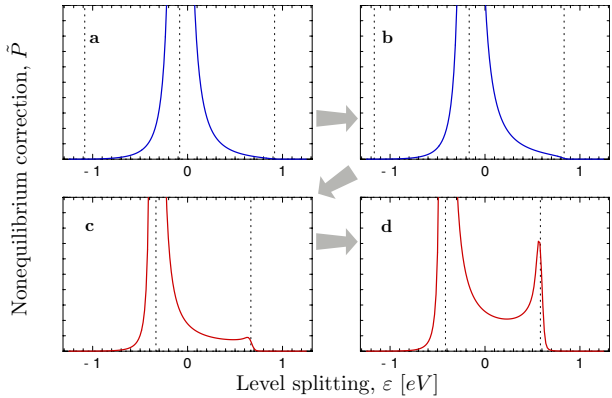}
\caption{{\bf The function $\tilde P(\varepsilon)$ that contains the effect of the bias voltage $V$.} 
As explained in the text, $\check{s}_2^\dagger \check{s}_1$ was parameterized
as in Eq.\ (\ref{eqs2}). A value $\phi=\tfrac{\pi}{9}$ is used througout. The values of $\theta$ in ({\bf a}), ({\bf b}), ({\bf c}) and ({\bf d})
are respectively $\tfrac{\pi}{6}$, $\tfrac{\pi}{3}$, $\tfrac{2\pi}{3}$ and $\tfrac{5\pi}{6}$. When $\theta<\pi/2$, then
$\tilde P(\varepsilon)$ has a fairly symmetric peak centered at $-eV\theta/2\pi$. The tails of this peak vanish at 
$\varepsilon\simeq(-\theta/2\pi\pm1)eV$. When $\theta>\pi/2$, there are two asymmetric peaks at $-eV\theta/2\pi$ and $(1-\theta/2\pi)eV$.
The value of $\tilde P(\varepsilon)$ is significantly larger for $\varepsilon\in[-eV\theta/2\pi,(1-\theta/2\pi)eV]$ than outside this
interval.\label{sfig1}}
\end{center}
\end{figure}
In the main text we confined our attention to the one parameter family of scattering matrices
\begin{equation}
\check{s}_2^\dagger \check{s}_1=\left(\begin{array}{cc}\cos\phi&i\sin\phi\\i\sin\phi&\cos\phi\end{array}\right).\label{eqs1}
\end{equation}
For this choice, $\tilde P(\tau)$ is a real function of time. For $\theta<\pi/2$ its fluctuations are associated with energies $\sim\pm eV$ due
to the transfer of eigenvalues from $1$ to $\cos\phi$ at a rate of one per $h/eV$. For $\phi>\pi/2$ however, $\cos\phi$ is
negative and two eigenvalues have to be transfered before the sign of $\tilde P(\tau)$ returns to its initial value. The
period of fluctuantions of $\tilde P(\tau)$ doubles and becomes associated with energies $\pm eV/2$. Because $\tilde P(\tau)$ is real,
the fluctuations with positive and negative energies are equal: $\tilde P(\varepsilon)=\tilde P(-\varepsilon)$. This translates into the
following feature of the probability $p_2$ to find the qubit in state $\left|2\right>$. For $\phi<\pi/2$, $p_2(\varepsilon)$ changes
from $1$ to $0$ in an energy interval of length $2 eV$. For $\phi>\pi/2$, this interval shrinks to $eV$. The boundry of
the interval is defined more sharply the closer $\phi$ is to $0$ or $\pi$. 
The shrinking from $2eV$ to $eV$ of the interval 
in which $p_2$ varies significantly is explained in terms of charge fractionalization: For $\phi>\pi/2$ the excitations in the
QPC transmit half the charge of an electron 
so that the energy that the qubit can absorb from the QPC changes from $eV$ to $eV/2$.
\begin{figure}[b]
\begin{center}
\includegraphics[width=.94 \columnwidth]{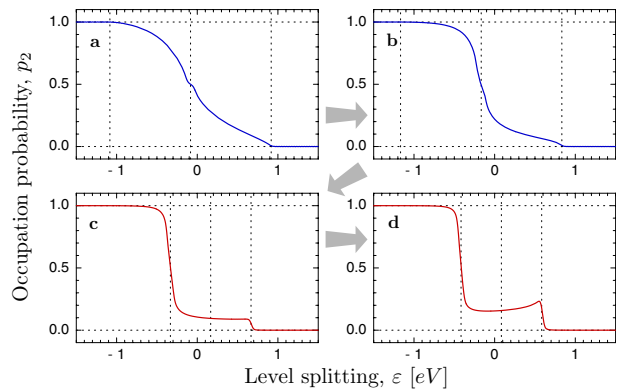}
\caption{{\bf The probability $p_2(\varepsilon)$.} $\check{s}_2^\dagger \check{s}_1$ is chosen as in Fig.\ (\ref{sfig1}):
A value $\phi=\tfrac{\pi}{9}$ is used througout. The values of $\theta$ in ({\bf a}), ({\bf b}), ({\bf c}) and ({\bf d})
are respectively $\tfrac{\pi}{6}$, $\tfrac{\pi}{3}$, $\tfrac{2\pi}{3}$ and $\tfrac{5\pi}{6}$. When $\theta<\pi/2$, the occupation probability
$p_2$ is significantly different from its asymptotic values $0$ and $1$ in an $\varepsilon$ interval of $2eV$.
When $\theta>\pi/2$, this interval shrinks to $eV$. The boundaries of the interval are more sharply defined the closer
$\theta$ is to $\pi/2$. The shrinking of the interval corresponds to a cross-over in the QPC from excitations that
transmit charge $e$ to excitations that transmit charge $e/2$.
\label{sfig2}}
\end{center}
\end{figure}

Since the QPC scattering matrices contain parameters that are not under experimental control, it is relevant to ask how the results 
are altered when a more general choice
\begin{equation}
\check{s}_2^\dagger \check{s}_1=\left(\begin{array}{cc}e^{-i\theta}\cos\phi&i\sin\phi\\i\sin\phi&e^{i\theta}\cos\phi\end{array}\right)\label{eqs2}
\end{equation}
with $\phi\in[-\tfrac{\pi}{2},\frac{\pi}{2}]$ and $\theta\in[0,\pi]$
is made for the scattering matrices. With this choice, eigenvalues travel from $1$ to $e^{i\theta}\cos\phi$ at a rate of one
per $2\pi/eV$. This means that the period doubling of $\tilde P(\tau)$ no longer takes place. The phase of $\tilde P(\tau)$ does not
return to its original value after the transfer of two eigenvalues. Rather, one expects fluctuations associated with an energy
$(n-\frac{\theta}{2\pi})eV,\ n=0,\pm 1,\pm 2,\ldots$ Because $\tilde P(\tau)$ is no longer real, positive and negative frequencies
don't contribute equally. However, while the eigenvalue trajectories lie close to the real line, one can expect results similar
to those obtained for real $\tilde P(\tau)$.
We obtained numerical results for four scattering matrices of the form (\ref{eqs2}). We chose $\theta=\tfrac{1}{6}\pi,\, \tfrac{1}{3}\pi,
\tfrac{2}{3}\pi$ and $\tfrac{5}{6}\pi$. To sharpen abrupt features we chose $\phi=\pi/9$ so that the exponential
decay of $\tilde P(\tau)$ is associated with a long decoherence time: $\simeq0.06\hbar/eV$. As depicted in Fig.\ (\ref{sfig1}), we
found $\tilde P(\varepsilon)$ to behave as follows. For $\theta$ close to zero, $\tilde P(\varepsilon)$ consists of
one peak situated at $\varepsilon=-\frac{\theta}{2\pi}eV$. The tails of this peak vanish at $\varepsilon=\left(\pm 1-\frac{\theta}{2\pi}\right)eV$.
The closer to zero that $\theta$ is taken, the more abrupt this behavior of the tails become. As $\theta$ is increased, a second peak
starts appearing at $\varepsilon=\left(1-\frac{\theta}{2\pi}\right)eV$. When $\theta=\pi-0^+$, the height (and width) of this peak exactly equals that of the 
peak at $-\frac{\theta}{2\pi}eV$. In the interval $\varepsilon\in\left[-\frac{\theta}{2\pi}eV,\left(1-\frac{\theta}{2\pi}\right)eV\right]$
that is bounded by the peaks, $\tilde P(\tau)$ is significantly larger than in the region outside the peaks.
\begin{figure}[b]
\begin{center}
\includegraphics[width=.94 \columnwidth]{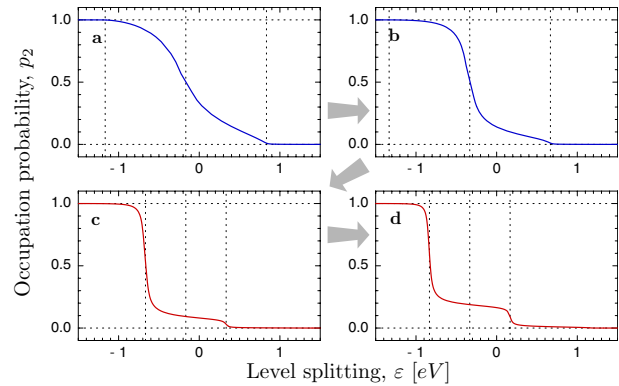}
\caption{{\bf The probability $p_2(\varepsilon)$ with spin included.} $\check{s}_2^\dagger \check{s}_1$ is chosen as in Fig.\ (\ref{sfig1})
and (\ref{sfig2}):
A value $\phi=\tfrac{\pi}{9}$ is used througout. The values of $\theta$ in ({\bf a}), ({\bf b}), ({\bf c}) and ({\bf d})
are respectively $\tfrac{\pi}{6}$, $\tfrac{\pi}{3}$, $\tfrac{2\pi}{3}$ and $\tfrac{5\pi}{6}$. Fractional charge
features are still clearly visible for $\theta>\pi/2$.\label{sfig3}}
\end{center}
\end{figure}
This behavior of $\tilde P(\varepsilon)$ translates into the occupation probabilities $p_2(\varepsilon)$ depicted in Fig.\ (\ref{sfig2}).
For $\theta<\pi/2$, $p_2(\varepsilon)$ still changes from unity to zero in an interval of length $2eV$ manifesting excitations with 
charge $e$ while for $\theta>\pi/2$ the interval
shrinks to $eV$, indication half-integer charge. 
The closer $\theta$ moves to $0$ or $\pi$, the sharper the interval becomes defined. We therefore conclude that the
fractional charge phenomenon in the QPC is not confined to the special choice (\ref{eqs1}) of scattering matrices.
\section{Inclusion of spin}
Up to this point we have considered spinless electrons in the QPC. In this Appendix we investigate the effect of including spin.
We still take the interaction between the QPC and the qubit to be spin independent. However, the mere existence of a spin degree
of freedom for QPC electrons doubles the dimension of channel space. The narrowest QPC now has
two channels in stead of one and
$\tilde P_{s=\tfrac{1}{2}}(\tau)=\tilde P_{s=0}(\tau)^2$, i.e. the determinant $\tilde P_{s=\tfrac{1}{2}}(\tau)$ with spin
included is the square of the determinant $\tilde P_{s=0}(\tau)$ without spin. For real determinants, squaring kills the phase.
This means that the observed period doubling for the parametrization of Eq.\ (\ref{eqs1})  disappears and with it the half integer charge features
of $p_2$. Physically, it could be that two charge $e/2$ excitations are transmitted through the QPC simultaneously. However, fractional charge is
saved by the fact
that, for $\theta\not=0$, $\tilde P_{s=0}(\varepsilon)$ has two peaks with different heights.
Suppose the relative peak heights are $A$ and $1-A$, i.e. 
\begin{equation}
\tilde P_{s=0}(\tau)\sim (1-A)e^{i\tfrac{\theta}{2\pi}eV\tau}+Ae^{-(1-\tfrac{\theta}{2\pi})eV\tau}
\end{equation}
where $A$ is a real number between $0$ and $\tfrac{1}{2}$. ($A=0$ corresponds to $\theta=0$ while $A=\tfrac{1}{2}$ corresponds to $\theta=\pi$.) 
It follows that $P_{s=\tfrac{1}{2}}(\varepsilon)$ has three peaks
at 
\begin{enumerate}
\item{$\varepsilon=-2\frac{\theta}{2\pi}eV$ with height $(1-A)^2$,}
\item{$\varepsilon=\left(1-2\frac{\theta}{2\pi}\right)eV$ with height $2A(1-A)$}
\item{and $\varepsilon=\left(2-2\frac{\theta}{2\pi}\right)eV$ with height $A^2$}
\end{enumerate}
As long as $A$ is small, i.e. $\theta$ is not too close to $\pi$, the first two peaks will dominate the third, and a signature of fractional
charge may still be observable in $p_2(\varepsilon)$. Fig.\ (\ref{sfig3}), contains $p_2$ calculated for the same scattering matrices as in
Fig. (\ref{sfig2}), but with spin included. The cases when $\theta=\tfrac{2}{3}\pi$ and $\theta=\tfrac{5}{6}\pi$ still contain clear half-integer
charge features. For $\theta$ very close to $\pi$ (not shown) these features disappear.


\begin{thebibliography}{99}
\bibitem{vWee88}{B.\ J.\ Van Wees {\em et al.},\ Phys.\ Rev.\ Lett.\ {\bf 60}\ 848\ (1988).}
\bibitem{But90}{M.\ B\"uttiker,\ Phys.\ Rev.\ B\ {\bf 41} 7906\ (1990).}
\bibitem{Lev96}{L.\ S.\ Levitov,\ H.\ Lee and  G.\ H.\ Lesovik,\ J.\ Math.\ Phys.\ {\bf 37}\ 4845\ (1996).}
\bibitem{Been03}{C.\ W.\ J.\ Beenakker,\ C.\ Emary and M.\ Kindermann,\ Phys.\ Rev.\ Lett.\ {\bf 91}, Art.\ No.\ 147901\ (2003).}
\bibitem{Elz03}{J.\ M.\ Elzerman,\ Phys.\ Rev.\ B\ {\bf 67}\ Art.\ No.\ 161308(R) (2003).}
\bibitem{Pet04}{J.\ R.\ Petta {\em eta.}\ Phys.\ Rev.\ Lett.\ {\bf 93}\ Art.\ No.\ 186802 (2004).} 
\bibitem{Ale97}{I.\ L.\ Aleiner,\ I.\ L., N.\ S.\ Wingreen and Y.\ Meir,\ Phys.\ Rev.\ Lett.\ {\bf 79} 3740\ (1997).}
\bibitem{Levi97}{Y.\ Levinson,\ Europhys.\ Lett.\ {\bf 39} 299\ (1997)}
\bibitem{Onac06}{E.\ Onac,\ Phys.\ Rev.\ Lett.\ {\bf 96}\ Art.\ No.\ 176601 (2006).}
\bibitem{Tob06}{J.\ Tobiska,\ J.\  Danon,\ I.\ Snyman and Y.\ V.\ Nazarov,\ Phys.\ Rev.\ Lett.\ {\bf 96}\ Art.\ No.\ 096801\ (2006).}
\bibitem{Lau83}{R.\ B.\ Laughlin,\ Phys.\ Rev.\ Lett.\ {\bf 50} 1395\ (1983).}
\bibitem{Jac76}{R.\ Jackiw and C.\ Rebbi,\ Phys. Rev. D {\bf 13} 3398\ (1976).}
\bibitem{Sut90}{B.\ Sutherland and B.\ S.\ Shastry,\ Phys.\ Rev.\ Lett.\ {\bf 65}, 1833\ (1990).}
\bibitem{Naz03}{Y.\ V.\ Nazarov and M.\ Kindermann,\ Euro.\ Phys.\ J.\ B\ {\bf 35}, 413\ (2003).} 
\bibitem{Aba04}{D.\ A.\ Abanin and L.\ S.\ Levitov,\ Phys. Rev. Lett. {\bf 93}\ Art.\ No.\ 126802\ (2004).}
\bibitem{Aba05}{D.\ A.\ Abanin and L.\ S.\ Levitov,\ Phys. Rev. Lett. {\bf 94}\ Art.\ No.\ 186803\ (2005).}
\bibitem{And67}{P.\ W.\ Anderson,\ Phys.\ Rev.\ Lett.\ {\bf 24} 1049 (1967).}
\bibitem{Mah67}{G.\ D.\ Mahan,\ Phys.\ Rev.\ {\bf 163}\ 612\ (1967).}
\bibitem{Noz69}{P. Nozi\`eres\ and C.\ T.\  De\ Dominicic,\ Phys.\ Rev.\ {\bf 178}\ 1097-1107\ (1969).}
\bibitem{Mat92}{K.\ A.\ Matveev\ and A.\ I.\ Larkin,\ Phys.\ Rev.\ B\ {\bf 46} 15337-15347\ (1992).}
\bibitem{Yam82}{K.\ Yamada and K.\ Yosida,\ Prog.\ Th.\ Phys.\ {\bf 68} 1504 (1982).}
\end{thebibliography}
\end{document}